\documentclass[preprint,showpacs,showkeys,preprintnumbers,amsmath,amssymb]{revtex4}

\usepackage{graphicx}
\usepackage{dcolumn}
\usepackage{bm}

\begin{document}


\title{Nonlinear stochastic models of $1/f$ noise and power-law distributions}

\author{Bronislovas Kaulakys}
 \email{kaulakys@itpa.lt}
\author{Julius Ruseckas}
\author{Vygintas Gontis}
\author{Miglius Alaburda}
\affiliation{
Institute of Theoretical Physics and Astronomy of Vilnius University,\\
A. Go\v{s}tauto 12, LT-01108 Vilnius, Lithuania
}

\date{\today}

\begin{abstract}
Starting from the developed generalized point process model of $1/f$ noise (B. Kaulakys et al, Phys. Rev. E 71 (2005) 
051105; cond-mat/0504025) we derive the nonlinear stochastic differential equations for the signal exhibiting $1/f^{\beta}$ noise and $1/x^{\lambda}$ distribution density of the signal intensity with different values of $\beta$ and $\lambda$. 
The processes with $1/f^{\beta}$ are demonstrated by the numerical solution of the derived equations with the 
appropriate restriction of the diffusion of the signal in some finite interval. The proposed consideration may be 
used for modeling and analysis of stochastic processes in different systems with the power-law distributions, long-range memory or with the elements of self-organization.
\end{abstract}


\maketitle

\section{Introduction}

The entirely uncorrelated in time stochastic signals exhibit white noise -- the power spectral density $S(f)$ as a 
function of the frequency $f$ is constant, while the Brownian motion of the signal intensity without correlations 
between increments results in $1/f^{2}$ and in the Lorentzian power spectra.
The widespread occurring signals and processes with $1/f$ spectrum (see, e.g. \cite{weismann,gilden,thurner,wong,KGA} 
and references therein) cannot be understood and modeled in such a way. 

``$1/f$ noise'' is a type of stochastic processes which power spectral density at low frequencies behaves like 
$S(f)\sim 1/f^{\beta}$, where the exponent $\beta$ is close to 1. 
In contrast to the Brownian motion and $1/f^{2}$ noise generated by the linear stochastic equation, simple systems 
of differential equations, even linear stochastic equations, generating signals with $1/f$ noise are not known.

Recently, starting from the simple point process model of $1/f$ noise \cite{KM,BK99}, we derived the nonlinear stochastic differential equation 
\begin{equation}
\frac{dx}{dt_{s}}=x^{4}+x^{5/2}\xi(t_{s})
\label{eq:1}
\end{equation} 
for the signal intensity $x$ generating processes with pure $1/f$ noise and the inverse cubic, $P(x)\sim 1/x^{3}$, 
distribution of the signal intensity \cite{KR}. 
Here $t_{s}$ is the scaled time and $\xi(t_{s})$ is a Gaussian white noise satisfying the standard condition 
\begin{equation}
\langle\xi(t_{s})\xi(t_{s}^{\prime})\rangle=\delta(t_{s}-t_{s}^{\prime})
\label{eq:2}
\end{equation} 
with the brackets $\langle\ldots\rangle$ denoting the averaging over the realizations of the process.

The aims of this paper are the derivation of the class of the stochastic nonlinear differential equations exhibiting 
$1/f^{\beta}$ noise and $1/x^{\lambda}$ distribution density of the signal with different values of the exponents
$\beta$ and $\lambda$ and the numerical demonstration of the proposed model for generation of the long-range fractal
processes.

\section{The model} 

We start from the point process as a sequence of correlated pulses or series of events 
\begin{equation}
I(t)=a\sum_{k}\delta(t-t_{k}). 
\label{eq:3}
\end{equation} 
Here $\delta(t)$ is the Dirac $\delta$-function and $a$ is a contribution to the signal or current of one pulse at the
time moment $t_{k}$. We consider the stochastic multiplicative process for the interevent time $\tau_{k}=t_{k+1}-t_{k}$
\cite{KGA,GK}
\begin{equation}
\tau_{k+1}=\tau_{k}+\gamma\tau_{k}^{2\mu-1}+\sigma\tau_{k}^{\mu}\varepsilon_{k},
\label{eq:4}
\end{equation}
where the (average) interevent time fluctuates due to the random perturbation by a sequence of uncorrelated normally 
distributed random variables $\{\varepsilon_{k}\}$ with zero expectation and unit variance, $\sigma$ denotes the 
standard deviation of this white noise and $\gamma\ll 1$ is a coefficient of the nonlinear damping.
It has been shown analytically and numerically \cite{KGA,GK} that the process \eqref{eq:3} and \eqref{eq:4} may
generate signals with the power-law distributions of the signal intensity and $1/f^{\beta}$ noise.

Transformation of Eq.~\eqref{eq:4} to the It\^{o} stochastic differential equation in $k$-space is 
\begin{equation}
\frac{d\tau_{k}}{dk}=\gamma\tau_{k}^{2\mu-1}+\sigma\tau_{k}^{\mu}\xi(k).
\label{eq:5}
\end{equation}
Transition from the occurrence number $k$ to the actual time $t$ according to the relation $dt=\tau_{k}dk$ yields
\begin{equation}
\frac{d\tau}{dt}=\gamma\tau^{2\mu-2}+\sigma\tau^{\mu-1/2}\xi(t).
\label{eq:6}
\end{equation}
The standard transformation \cite{gardiner} of the variable from $\tau_{k}$ to the averaged over the time interval $\tau_{k}$ intensity of the signal 
$x=a/\tau_{k}$ by analogy with Ref.~\cite{KR} yields the stochastic nonlinear differential equation 
\begin{equation}
\frac{dx}{dt}=(\sigma^{2}-\gamma)\frac{x^{4-2\mu}}{a^{3-2\mu}}+\frac{\sigma x^{5/2-\mu}}{a^{3/2-\mu}}\xi(t).
\label{eq:7}
\end{equation}
Introducing the scaled time
\begin{equation}
t_{s}=\frac{\sigma^{2}}{a^{3-2\mu}}t
\label{eq:8}
\end{equation}
and the new parameters
\begin{equation}
\eta=\frac{5}{2}-\mu, \quad \Gamma=1-\frac{\gamma}{\sigma^{2}}
\label{eq:9}
\end{equation}
we obtain the class of It\^{o} stochastic differential equations 
\begin{equation}
\frac{dx}{dt_{s}}=\Gamma x^{2\eta-1}+x^{\eta}\xi(t_{s}). 
\label{eq:10}
\end{equation}
Eqs.~\eqref{eq:10}, as far as it corresponds to the analyzed in Refs. \cite{KGA,GK} point process \eqref{eq:3}-\eqref{eq:4}, should generate the signals with the power-law distributions of the signal intensity, 
\begin{equation}
P(x)\sim\frac{1}{x^{\lambda}},\quad\lambda=2(\eta-\Gamma), 
\label{eq:11}
\end{equation}
and $1/f^{\beta}$ noise,
\begin{equation}
S(f)\sim\frac{1}{f^{\beta}},\quad\beta=2-\frac{2\Gamma+1}{2\eta-2}.
\label{eq:12}
\end{equation}

According to the general theory \cite{gardiner} the exponentially restricted diffusion with the distribution densities 
\begin{equation}
P(x)\sim\frac{1}{x^{\lambda}}\exp\left\{-\left(\frac{x_{\mathrm{min}}}{x}\right)^m
-\left(\frac{x}{x_{\mathrm{max}}}\right)^m\right\}\label{eq:11a}
\end{equation}
should generate the stochastic differential equations 
\begin{equation}
\frac{dx}{dt_{s}}=\frac{m}{2}\left(\frac{x_{\min}^{m}}{x^{m+1-2\eta}}-\frac{x^{m-1+2\eta}}{x_{\max}^{m}}\right)+
\Gamma x^{2\eta-1}+x^{\eta}\xi(t_{s}),
\label{eq:13}
\end{equation}
where $m$ is some parameter.

\section{Numerical analysis}

\begin{figure*}
\begin{center}
\includegraphics[width=.5\textwidth]{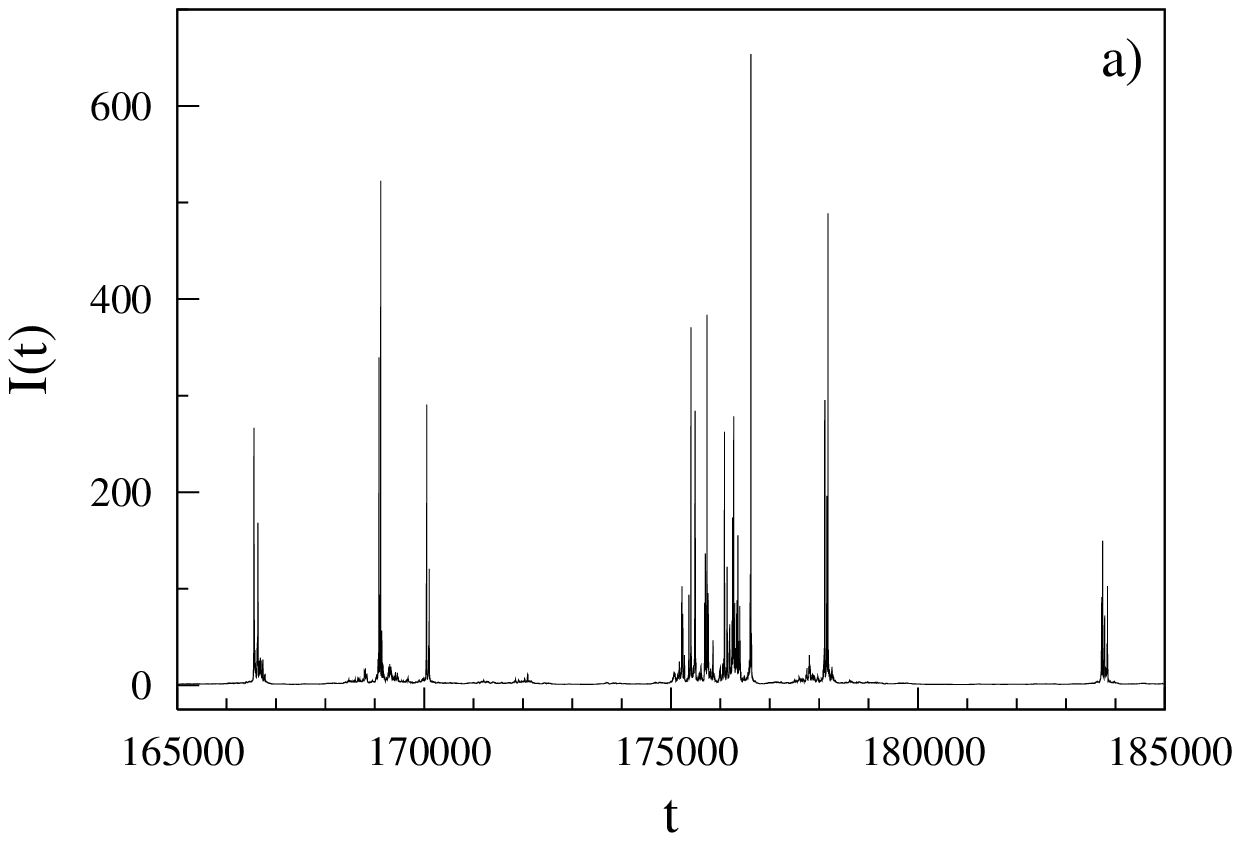}
\hspace{-10pt}
\includegraphics[width=.5\textwidth]{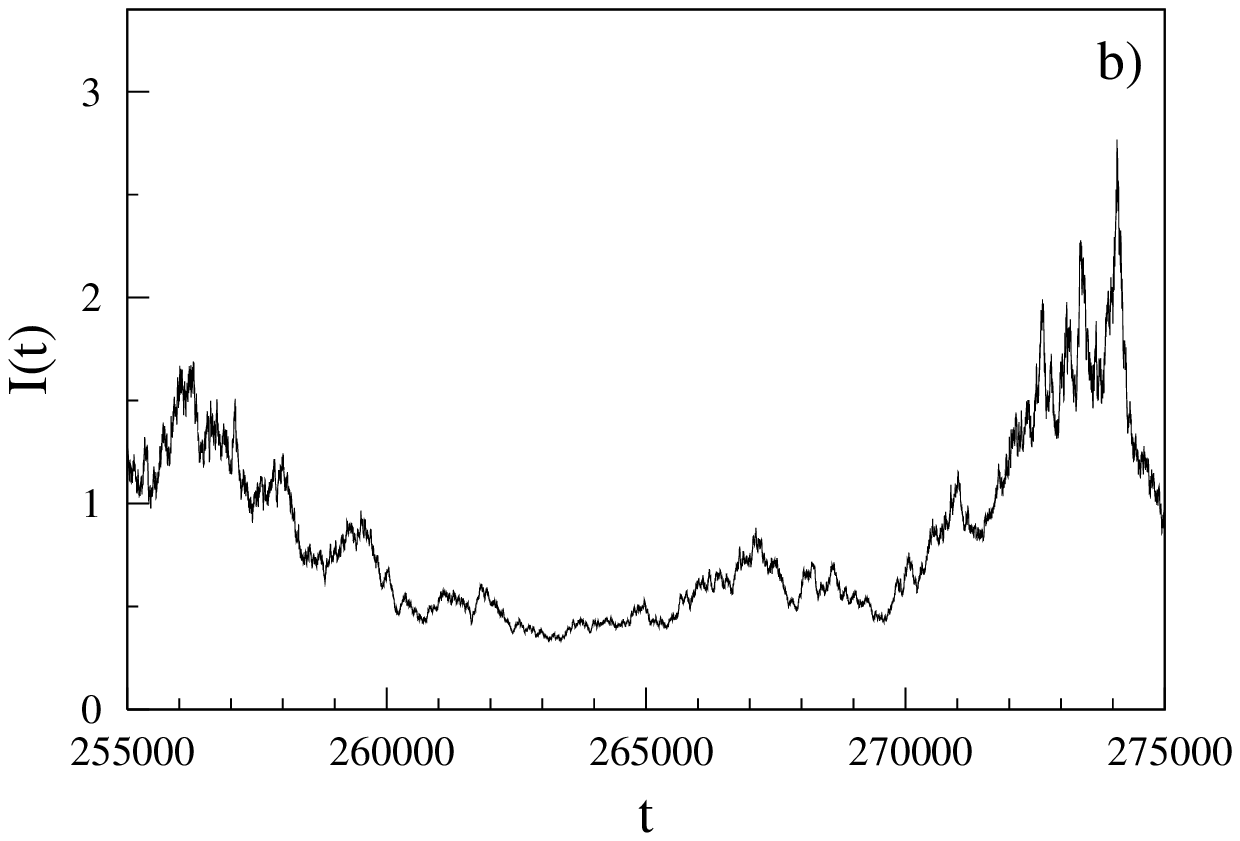}
\end{center}
\par
\vspace{-10pt}
\caption{Typical examples of the solutions of Eqs.~\eqref{eq:13} and \eqref{eq:14}: a) with the parameters 
$\Gamma=0.75$, $\eta=2$, and $\kappa=0.1$ and b) with the parameters $\Gamma=-0.2$, $\eta=1.5$, and $\kappa=0.01$.}
\label{fig:1}
\end{figure*}

For the numerical solution of Eq.~\eqref{eq:13} we can take the integration steps from the equation 
$x_{i}^{\eta}\sqrt{h_{i}}=\kappa x_{i}$, with $\kappa\ll 1$ being a small parameter. This corresponds to the case when the change of the variable $x_i$ in one step is proportional to the value of the variable. 
As a result, we obtain the system of equations 
\begin{gather}
x_{i+1}=x_{i}+\kappa^{2}x_{i}\left[\Gamma+\frac{m}{2}\left(\frac{x_{\min}^{m}}{x_{i}^{m}}-
\frac{x_{i}^{m}}{x_{\max}^{m}}\right)\right]+\kappa x_{i}\varepsilon_{i}, \notag \\
t_{i+1}=t_{i}+\frac{\kappa^{2}}{x_{i}^{2\eta-2}}.
\label{eq:14}
\end{gather}

\begin{figure*}
\begin{center}
\includegraphics[width=.5\textwidth]{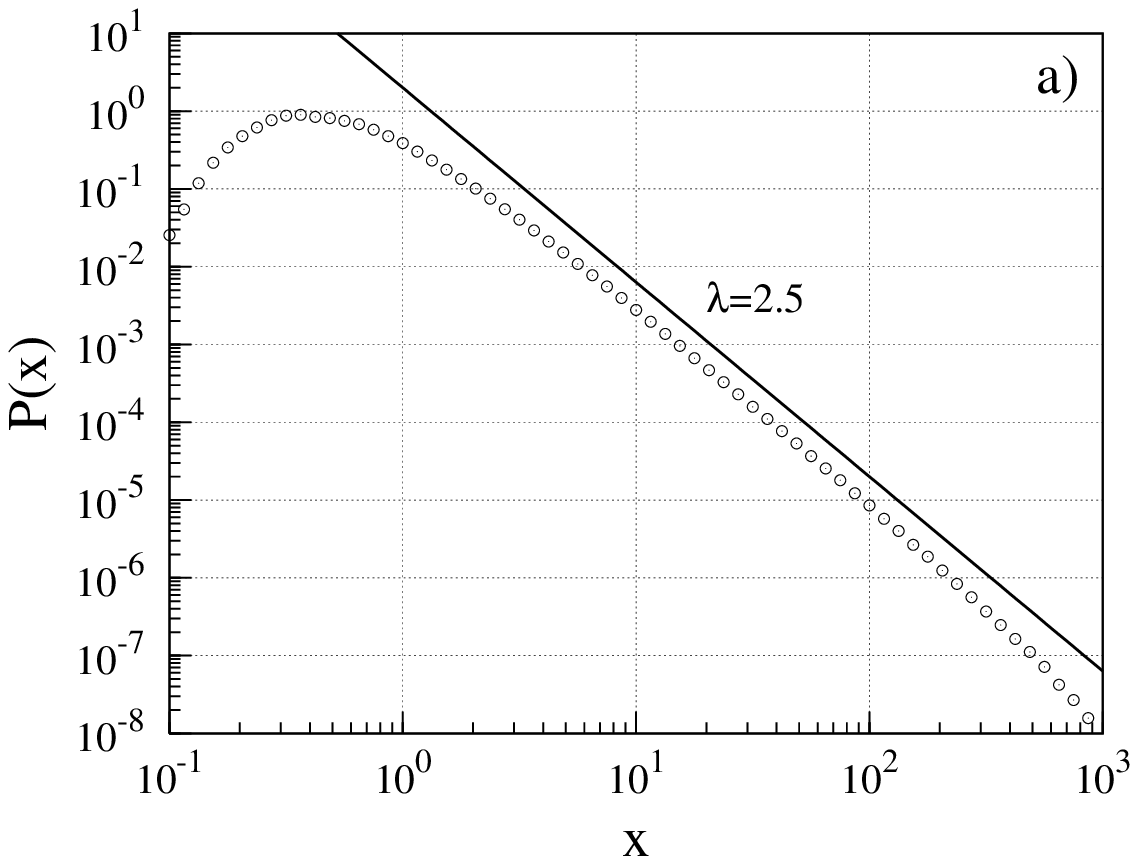}
\hspace{-10pt}
\includegraphics[width=.5\textwidth]{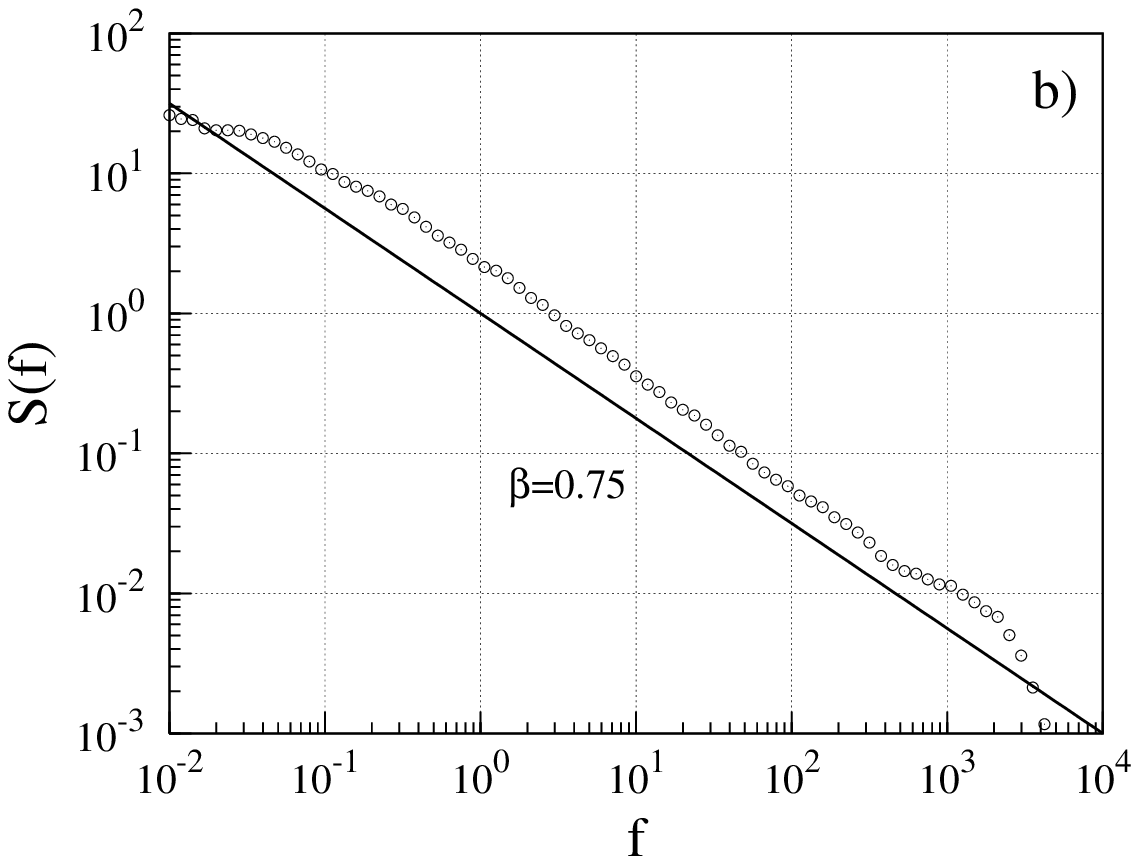}\\
\vspace{10pt}
\includegraphics[width=.5\textwidth]{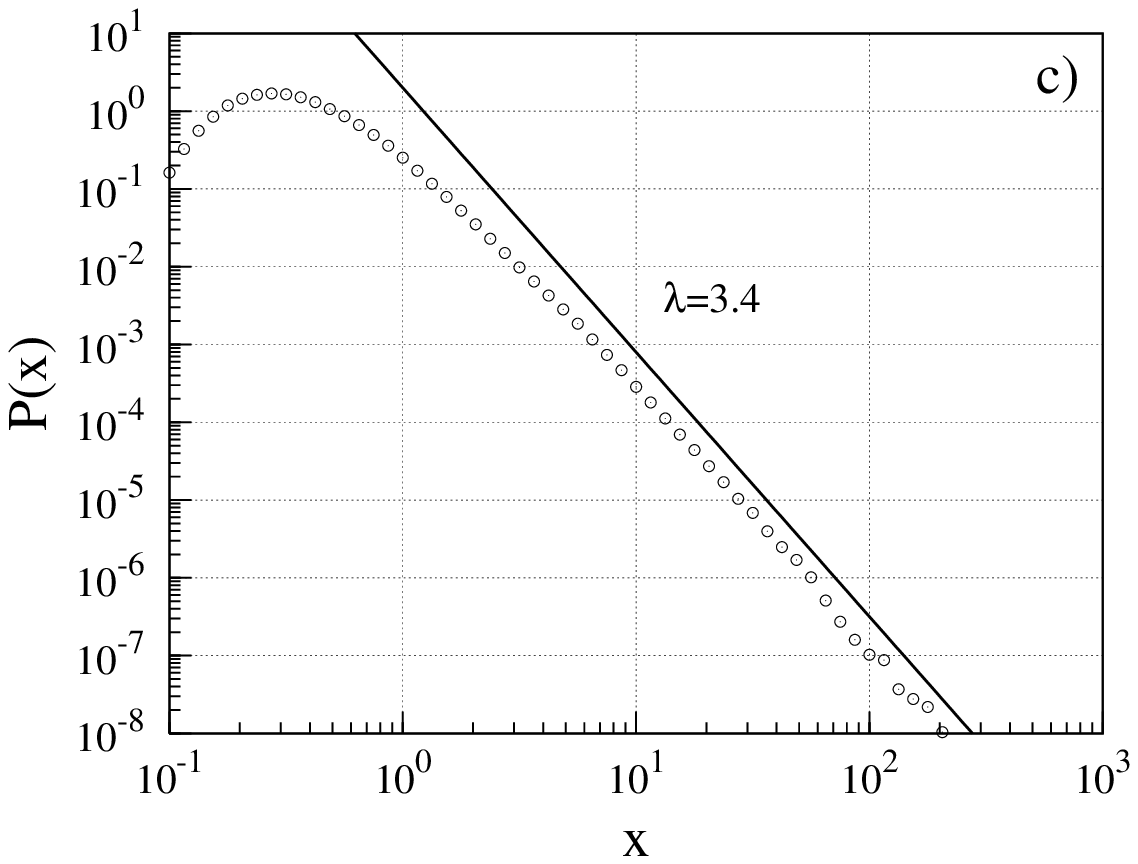}
\hspace{-10pt}
\includegraphics[width=.5\textwidth]{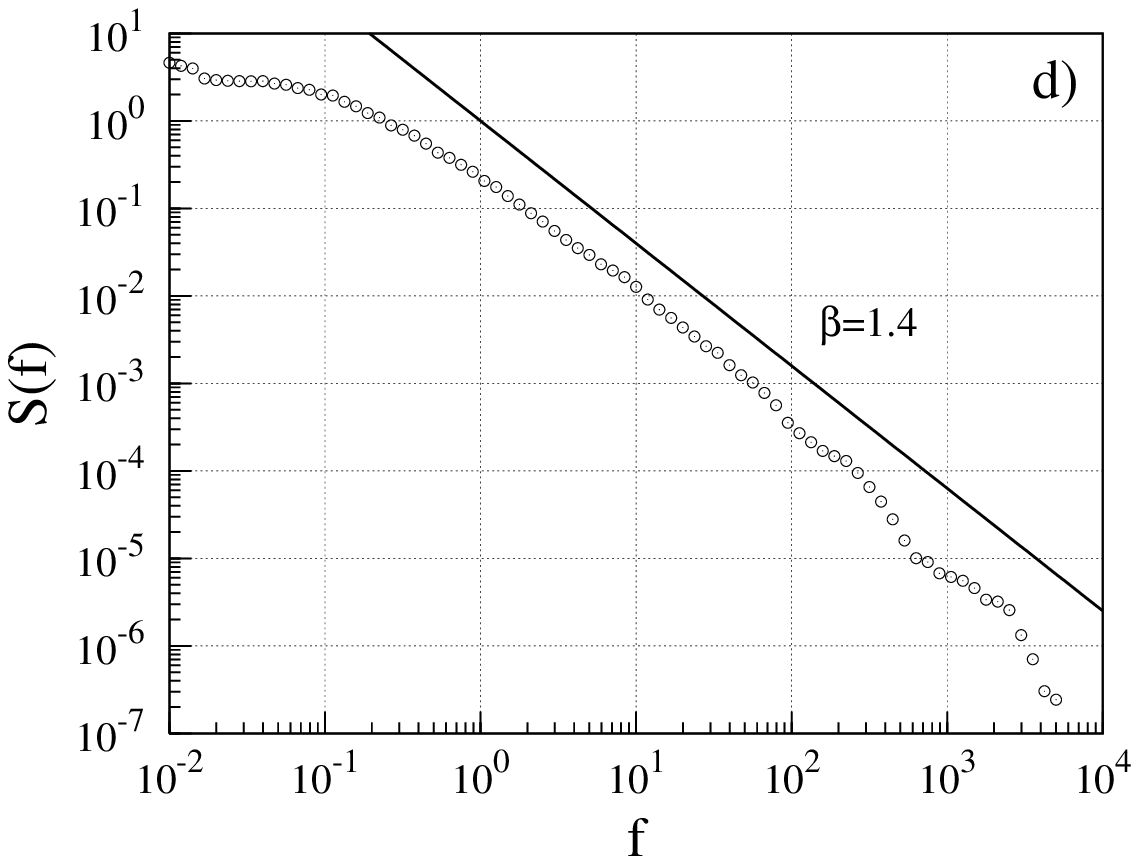}
\end{center}
\par
\vspace{-10pt}
\caption{a) Numerically simulated distribution density of the variable $x$ according to Eqs.~\eqref{eq:14}, 
open circles, compared with the expected distribution density \eqref{eq:11}, solid line; 
and b) power spectral density, obtained from the numerical solution of Eq.~\eqref{eq:14}, open circles. 
Solid line in b) represents the analytical power spectral density slope \eqref{eq:12}.
Parameters used are $x_{\min}=1$, $x_{\max}=10^{3}$, $m=1$, $\Gamma=0.75$, $\eta=2$, and $\kappa=0.1$.
The signal was calculated from $N_{x}=10^{6}$ points and averaged over $n=100$ realizations. 
c) and d) represent the distribution density and the power spectral density, respectively, with the parameters 
$m=1$, $\Gamma=-0.2$, $\eta=1.5$, and $\kappa=0.01$.}
\label{fig:2}
\end{figure*}
In figure~\ref{fig:1} the typical examples of the signals as solutions of Eqs.~\eqref{eq:13} and \eqref{eq:14} are 
shown. 
The distribution densities $P(x)$ of the variable $x$, obtained by the numerical simulation of Eq.~\eqref{eq:14}, are 
shown in figures \ref{fig:2} a) and c). The power spectral densities $S(f)$ are shown in figures \ref{fig:2} b) and d). 
Numerical simulation of distribution densities and power spectral densities are in good agreement with approximate 
expressions, Eq.~\eqref{eq:11} and Eq.~\eqref{eq:12}, respectively. 

\section{Conclusions}

We derived and analyzed a class of stochastic nonlinear differential equations for the signal exhibiting $1/f^{\beta}$
noise and $1/x^{\lambda}$ distribution density of the signal in any desirable wide range of frequency and of the signal intensity.
The proposed technique may be used for modeling of the stochastic processes in different systems (e.g., in financial 
systems \cite{GK,GK2,wang} and the Internet \cite{field,GKR}) with the power-law statistics of the signal characteristics.

\begin{acknowledgments}
The support by the Lithuanian State Science and Studies Foundation is acknowledged.
\end{acknowledgments}


\begin{thebibliography}{0}
\bibitem{weismann}
M.~B.~Weissman, Rev. Mod. Phys. 60 (1988) 537.
\bibitem{gilden}
D.~L.~Gilden, T.~Thornton, M.~W.~Mallon, Science 267 (1995) 1837.
\bibitem{thurner}
S.~Thurner, et al., Fractals 5 (1997) 565.
\bibitem{wong}
H.~Wong, Microelectron. Reliab. 43 (2003) 585.
\bibitem{KGA}
B.~Kaulakys, V.~Gontis, M.~Alaburda, Phys. Rev. E 71 (2005) 051105; cond-mat/0504025. 
\bibitem{KM}
B.~Kaulakys, T.~Me\v{s}kauskas, Phys. Rev. E 58 (1998) 7013; adap-org/9812003. 
\bibitem{BK99}
B.~Kaulakys, Phys. Lett. A 257 (1999) 37; adap-org/9806004; adap-org/9907008. 
\bibitem{KR}
B.~Kaulakys, J.~Ruseckas, Phys. Rev. E 70 (2004) 020101(R); cond-mat/0408507. 
\bibitem{GK}
V.~Gontis, B.~Kaulakys, Physica A 343 (2004) 505; cond-mat/0303089. 
\bibitem{gardiner}
C.~W.~Gardiner, Handbook of Stochastic Methods for Physics, Chemistry and Natural Science, Springer, Berlin, 1985. 
\bibitem{GK2}
V.~Gontis, B.~Kaulakys, Physica A 344 (2004) 128; cond-mat/0412723. 
\bibitem{wang}
S.~J.~Wang, C.~S.~Zhang, Physica A 354 (2005) 496. 
\bibitem{field}
A.~J.~Field, U.~Harder, P.~G.~Harrison, IEE Proceedings-Communications 151 (2004) 355. 
\bibitem{GKR}
V.~Gontis, B.~Kaulakys, J.~Ruseckas, AIP Conf. Proceed. 776 (2005) 144; cs.NI/0508131. 
\end{thebibliography}
\end{document}